\documentstyle[aps,preprint,tighten,psfig]{revtex}
\newcommand{\beq}    {\begin{equation}}
\newcommand{\eeq}    {\end{equation}}
\newcommand{\beqarr} {\begin{eqnarray}}
\newcommand{\eeqarr} {\end{eqnarray}}
\newcommand{\barr}   {\begin{array}}
\newcommand{\earr}   {\end{array}}

\newcommand{\lsim}{\mathrel{\mathop{\kern 0pt \rlap
  {\raise.2ex\hbox{$<$}}}
  \lower.9ex\hbox{\kern-.190em $\sim$}}}
\newcommand{\gsim}{\mathrel{\mathop{\kern 0pt \rlap
  {\raise.2ex\hbox{$>$}}}
  \lower.9ex\hbox{\kern-.190em $\sim$}}}

\begin{document}

\preprint{
\begin{tabular}{r}
DFTT 42/99 \\
FTUV/99--60, IFIC/99--63
\end{tabular}
}

\title{
Implications for relic neutralinos of the
 theoretical\\ uncertainties in the neutralino--nucleon
 cross--section}

\author{\bf 
A. Bottino$^{\mbox{a}}$
\footnote{E--mail: bottino@to.infn.it, donato@lapp.in2p3.fr, 
fornengo@flamenco.ific.uv.es, \\
\phantom{E--mail:~~~} scopel@posta.unizar.es},
F. Donato$^{\mbox{b}}$, 
N. Fornengo$^{\mbox{c}}$,
S. Scopel$^{\mbox{d}}$\footnote[4]{INFN Post--doctoral Fellow}
\vspace{6mm}
}

\address{
\begin{tabular}{c}
$^{\mbox{a}}$
Dipartimento di Fisica Teorica, Universit\`a di Torino \\
and INFN, Sez. di Torino, Via P. Giuria 1, I--10125 Torino, Italy\\
$^{\mbox{b}}$
Laboratoire de Physique  Th\'eorique LAPTH, B.P. 110, F--74941\\
Annecy--le--Vieux Cedex, France \\
and INFN, Sez. di Torino, Via P. Giuria 1, I--10125 Torino, Italy\\
$^{\mbox{c}}$
Instituto de F\'{\i}sica Corpuscular -- C.S.I.C. --
Departamento de F\'{\i}sica Te\`orica, \\
Universitat de Val\`encia, E-46100 Burjassot, Val\`encia, Spain \\
$^{\mbox{d}}$ 
Instituto de F\'\i sica Nuclear y Altas Energ\'\i as, 
Facultad de Ciencias, \\
Universidad de Zaragoza, Plaza de San Francisco s/n, E--50009 Zaragoza, Spain
\end{tabular}
}


\maketitle

\begin{abstract}
We discuss the effect induced on the neutralino--nucleon 
cross--section by the present uncertainties 
in the values of the quark masses and of the
quark scalar densities in the nucleon. 
We examine the implications of this aspect on the 
determination of the neutralino 
cosmological properties,
as derived from measurements of WIMP direct detection. We show that,  
within current theoretical uncertainties, the DAMA annual modulation 
data are compatible with a neutralino as a major dark matter 
component, to an extent which is even larger than the one previously 
derived. We also comment on implications of the mentioned uncertainties 
for experiments of indirect dark matter detection.   
\end{abstract}

\section{Introduction}

  The sensitivities of the experiments of direct search for WIMPs have  
remarkably improved in recent years, allowing now the exploration of 
sizeable regions of the physical parameter space of specific particle 
candidates for dark matter. This is the case of the neutralino \cite{bf}, 
for which 
some direct detection experiments are already capable of investigating  
significant features in domains of the parameter space which are also 
under current exploration at LEP2.  
 
  The signal searched for in experiments of WIMP direct detection is a 
convolution of the WIMP velocity distribution in the halo with the quantity 
$\rho_{\chi} \sigma_{el}$, where $\rho_{\chi}$ is the local 
WIMP matter density \cite{den}  and $\sigma_{el}$ is the WIMP--nucleus 
 elastic cross--section. 
Under the assumption that the WIMP has the two following properties: 
i) its  cross-section with matter is dominated by coherent effects, 
ii) its (spin-independent) couplings  are essentially the same for protons 
as for neutrons, then one can straightforwardly factor out   
a scalar (spin-independent)  WIMP--{\it nucleon} cross--section,   
$\sigma_{\rm scalar}^{\rm (nucleon)}$, 
in $\sigma_{el}$. 
In this instance,  the information derivable from any experiment of WIMP direct 
search may be directly formulated in terms of the quantity 
$\rho_{\chi} \sigma_{\rm scalar}^{\rm (nucleon)}$, and the sensitivities of various 
experiments  may be easily compared among each other 
\cite{bdmsbi}. Properties i) and ii) are actually satisfied almost everywhere
in the supersymmetric parameter space \cite{vda}, when the WIMP candidate
is a neutralino, which is the case explicitly discussed in the
present paper. 

Some of the WIMP direct search experiments are already sensitive, 
or are on the verge of becoming sensitive,   to the range 
$\rho_{\chi}^{0.3} \sigma_{\rm scalar}^{\rm (nucleon)} \sim$ a few $\cdot$  
$10^{-9} \div 1 \cdot 10^{-8}$ nbarn, where  
$\rho_{\chi}^{0.3}$ denotes the local density normalized to the standard 
value of 0.3 GeV cm$^{-3}$.  This goal has already been achieved by the 
DAMA experiment 
\cite{dama1}, which has  reported the indication of an annual 
modulation effect in its counting rate, compatible with values of the 
quantity $\rho_{\chi}^{0.3} \sigma_{\rm scalar}^{\rm (nucleon)}$ in the range 

\beq \label{eq:ranger}
3 \cdot 10^{-9} \; {\rm nbarn}\lsim \rho_{\chi}^{0.3} \sigma_{\rm scalar}^{\rm (nucleon)} 
\lsim 1 \cdot 10^{-8}\; {\rm nbarn}, 
\eeq

\noindent for values of the WIMP mass, correlated 
with $\rho_{\chi}^{0.3} \sigma_{\rm scalar}^{\rm (nucleon)}$,
which  extend over the range 
 30 GeV $\lsim m_{\chi} \lsim$ 130 GeV \cite{bbbdfps}. 
The region in the  $m_{\chi}$ -- 
$\rho_{\chi}^{0.3} \sigma_{\rm scalar}^{\rm (nucleon)}$ plane, 
singled out by the DAMA experiment at  2--$\sigma$ C.L., is the one depicted 
in Fig. 2 of Ref. \cite{bbbdfps} and therein (and here) denoted by $R_m$. 
Let us also notice that, taking into account the uncertainties in the local 
total dark matter density: 
0.1 GeV cm$^{-3} \leq \rho_l \leq $ 0.7 GeV cm$^{-3}$\cite{turner1},
when one assumes that a single WIMP candidate 
saturates $\rho_l$, the range of Eq.(\ref{eq:ranger}) implies for 
$\sigma_{\rm scalar}^{\rm (nucleon)}$:

\beq \label{eq:ranges}
1 \cdot 10^{-9} \; {\rm nbarn}\lsim  \sigma_{\rm scalar}^{\rm (nucleon)} 
\lsim 3 \cdot 10^{-8}\; {\rm nbarn}, 
\eeq

\noindent
 Another  experiment of WIMP direct detection which is 
now entering the upper part of the  range of Eq.(\ref{eq:ranger}) for 
$\rho_{\chi}^{0.3}  \sigma_{\rm scalar}^{\rm (nucleon)}$ is the CDMS experiment 
 \cite{cdms}. 

Once a given range for $\rho_{\chi}  \sigma_{\rm scalar}^{\rm (nucleon)}$ is 
singled out by an experiment, what are the implications for specific 
particle candidates? We addressed this question in 
Refs. \cite{noi1234,bbbdfps} in the case of the DAMA data, under the 
assumption that the reported indication of an annual modulation is 
interpreted in terms of an effect due to relic neutralinos.  We derived a 
number of features for the neutralino cosmological properties, by selecting 
the supersymmetric configurations on the basis of the range of 
Eq.(\ref{eq:ranger}),
appropriately correlated with the range of $m_{\chi}$.  In particular, we 
proved that this range for $\rho_{\chi}^{0.3} \sigma_{\rm scalar}^{\rm (nucleon)}$ 
is compatible with a neutralino as a sizeable 
component of dark matter. In the derivation of this property, 
the role played by the correlation between $\sigma_{\rm scalar}^{\rm (nucleon)}$   
and the neutralino--neutralino annihilation cross--section, $\sigma_{ann}$,
is crucial. 
In fact, these two cross-sections are normally both increasing or decreasing 
functions of the supersymmetric parameters. Since 
the neutralino relic abundance $\Omega_{\chi} h^2$ 
($\Omega_{\chi}$ is the neutralino cosmological density
and $h$ is the Hubble parameter in units of
100 km s$^{-1}$ Mpc$^{-1}$) is roughly 
inversely proportional to $\sigma_{ann}$ \cite{bf},
lower bounds 
on $\sigma_{\rm scalar}^{\rm (nucleon)}$ entail upper bounds for the relic 
abundance. 
It is remarkable that, as mentioned above, the range of Eq.(1), 
singled out by the indication of annual 
modulation,  is compatible with a neutralino relic abundance of cosmological 
interest \cite{bbbdfps,noi1234}.   
 
From  the previous discussion it is clear that one of the crucial 
ingredients  
in the derivation of the neutralino relic abundance from the results of direct 
detection experiments is $\sigma_{\rm scalar}^{\rm (nucleon)}$. But, actually, how 
accurately can this quantity be evaluated at present?   
The point is that $\sigma_{\rm scalar}^{\rm (nucleon)}$ usually takes dominant 
contributions from interaction processes, where neutralinos and quarks 
(inside the nucleon) 
interact by exchange of Higgs particles or squarks, 
and the relevant 
couplings are still plagued  by sizeable uncertainties, related to
hadron physics, which have not yet received satisfactory answers. 
Indeed, the Higgs--quark--quark or squark--quark--neutralino
couplings involve the use of quark masses and quark 
scalar densities inside the nucleon, {\it i.e.} quantities which are 
subject to large uncertainties.  
  
We pointed out this problem in Ref. \cite{don}, where we showed that, by taking 
two different determinations \cite {tpcheng,gls} of the 
pion--nucleon sigma term, $\sigma_{\pi N}$,  and of the fractional strange 
content 
in the nucleon $y$ (see later on for the definitions of these two quantities),
 these uncertainties may affect $\sigma_{\rm scalar}^{\rm (nucleon)}$ approximately 
by a factor of 3.  This point, subsequently also recognized in Ref. \cite{jkg}, 
was usually overlooked in subsequent papers, though the afore mentioned 
uncertainties still persist. The common practice
prevailed of employing, for the relevant couplings, some standard values, 
which became consolidated more because of their reiterated use, rather
than because of confirmation by more refined theoretical analyses. 

Actually, the quantities $\sigma_{\pi N}$ and $y$ have recently 
been the object of 
various other calculations, based mainly on chiral perturbation theory and 
on QCD simulations on a lattice; however,  
the present situation, which is schematically revised in the next section, 
is still far from being clear. Thus, the Higgs--quark--quark and
squark--quark--neutralino couplings 
are still plagued by significant 
uncertainties, of the order of those pointed out in 
Ref. \cite{don}.
In the present paper we wish to take up  this problem again and, moreover, 
to address the question about how sensitive are 
the neutralino cosmological and astrophysical properties,
as derived from direct detection 
measurements, to the uncertainties which still affect the 
neutralino--nucleon cross--section. 
 
The theoretical framework employed here is the Minimal Supersymmetric 
extension of the Standard Model (MSSM), with the specifications given 
in Refs. \cite{noi1234,bbbdfps}, to which we refer for all relevant 
details. Our results will mainly be given in terms of scatter 
plots, obtained 
by scanning the supersymmetric parameter space over the grid defined in 
Refs. \cite{noi1234,bbbdfps}.
Of particular importance for the properties discussed in the 
next section is the role 
of the neutral Higgs bosons: the two CP--even ones, $h$ and $H$, and 
the CP--odd one, $A$. $h$ and $H$ are the main mediators of the 
coherent neutralino--nucleus cross--section, $A$ is important 
in the neutralino--neutralino annihilation channels. 
The neutralino is defined as the lowest--mass linear 
superposition of photino ($\tilde \gamma$),
zino ($\tilde Z$) and the two higgsino states
($\tilde H_1^{\circ}$, $\tilde H_2^{\circ}$)
\begin{equation}
\chi \equiv a_1 \tilde \gamma + a_2 \tilde Z + a_3 \tilde H_1^{\circ}  
+ a_4 \tilde H_2^{\circ} \, . 
\label{eq:neu}
\end{equation}

To classify the nature of the neutralino, we define a parameter 
$P \equiv a_1^2 + a_2^2$; hereafter the neutralino is called a {\it gaugino}, 
when $P > 0.9$, is called 
{\it mixed} when $0.1 \leq P \leq 0.9$ and a {\it higgsino} when $P < 0.1$. 
  
\section{Neutralino--nucleus elastic cross--section}

We recall that the neutralino--nucleon scalar cross--section is given by
\beq
\sigma_{\rm scalar}^{(\rm nucleon)} = \frac {8 G_F^2} {\pi} M_Z^2 m_{\rm red}^2 
\left[\frac{F_h I_h}{m_h^2}+\frac{F_H I_H}{m_H^2}+
\frac{M_Z}{2} \sum_q <N|\bar{q}q|N>
\sum_i P_{\tilde{q}_i} ( A_{\tilde{q}_i}^2- B_{\tilde{q}_i}^2) \right]^2, 
\label{eq:sigma}
\eeq
\noindent
where the two first terms inside the brackets refer to the diagrams with 
$h$-- and $H$--exchanges in the t--channel 
(the $A$--exchange diagram is strongly kinematically suppressed and then omitted
here) \cite{barbieri} and the third term refers 
to the graphs with squark--exchanges in the s-- and u--channels \cite{griest}.
The mass $m_{red}$ is the neutralino--nucleon
reduced mass. Here, for simplicity, we explicitly discuss only the
Higgs--mediated terms,
then we do not report the expressions for the squark propagator
$P_{\tilde{q}_i}$ and for the couplings $A_{\tilde{q}_i}$, $B_{\tilde{q}_i}$,
which may be found in Ref.\cite{noi1234}. However, arguments analogous
to the ones given below
hold also for the squark--exchange terms, which are actually included
in the numerical results reported in this paper.

The quantities $F_{h,H}$ and 
$I_{h,H}$ are defined as follows 
\beqarr
F_h &=& (-a_1 \sin \theta_W+a_2 \cos \theta_W) 
          (a_3 \sin \alpha + a_4 \cos \alpha) 
           \nonumber \\
           F_H &=& (-a_1 \sin \theta_W+a_2 \cos \theta_W) 
           (a_3 \cos \alpha - a_4 \sin \alpha) \nonumber \\
           I_{h,H}&=&\sum_q k_q^{h,H} m_q \langle N|\bar{q} q |N \rangle.
\label{eq:effe}
\eeqarr
\noindent The matrix elements $<N|\bar{q}q|N>$ are meant over the nucleonic
state.
The angle $\alpha$ rotates $H_1^{(0)}$ and $H_2^{(0)}$ into $h$ and $H$, and 
the coefficients $k_q^{h,H}$ are given by 

\beqarr
k_{u{\rm -type}}^h =   \cos\alpha / \sin\beta &,\;\;&  
k_{u{\rm -type}}^H = - \sin\alpha / \sin\beta\nonumber\\
k_{d{\rm -type}}^h = - \sin\alpha / \cos\beta &,\;\;& 
k_{d{\rm -type}}^H = - \cos\alpha / \cos\beta \label{eq:k}
\eeqarr
\noindent for the up--type and down--type quarks, respectively.
 
The quantities $ m_q \langle N|\bar{q} q |N \rangle $'s 
for the light quarks $u,d,s$ 
may conveniently be expressed in terms of the pion--nucleon sigma term
\beq
\sigma_{\pi N} = \frac{1}{2} (m_u + m_d) <N|\bar uu + \bar dd|N>,
\eeq
\noindent the fractional strange--quark content of the nucleon
\beq \label{eq:y}
y=2 \frac{<N|\bar ss|N>}
{<N|\bar uu+ \bar dd|N>},
\eeq
\noindent and the ratio $r=2 m_s/(m_u+m_d)$. 
In fact, assuming isospin invariance for quarks 
$u$ and $d$, one has
\beqarr
m_u <N|\bar uu|N>&\simeq& m_d <N|\bar dd|N> \simeq \frac{1}{2} \sigma_{\pi N}
\label{eq:condlight}\\
m_s <N|\bar ss|N>&\simeq& \frac{1}{2} r y \sigma_{\pi N}.\label{eq:condstrange}
\eeqarr 
For the heavy quarks $c$, $b$, $t$, using the heavy quark expansion\cite{svz},
one
derives
\beqarr
m_c <N|\bar cc|N>&\simeq& m_b <N|\bar bb|N> \simeq m_t <N|\bar tt|N> 
\simeq \nonumber\\
&\simeq& \frac{2}{27} \left [ m_N-(1+\frac{1}{2}ry)\sigma_{\pi N}\right],
\label{eq:condheavy}
\eeqarr
\noindent where $m_N$ is the nucleon mass. The quantities 
$I_{h,H}$ can then be rewritten as
\beq
I_{h,H} = k_{u{\rm -type}}^{h,H} g_u + k_{d{\rm -type}}^{h,H} g_d,
\label{eq:i}
\eeq
\noindent where
\beq
g_u = \frac {4} {27} (m_N + \frac {19}{8} \sigma_{\pi N} 
- \frac{1}{2}ry \sigma_{\pi N}),~~~~~g_d = \frac{2}{27} (m_N + \frac{23}{4} \sigma_{\pi N} 
+ \frac{25}{4} r y \sigma_{\pi N}). \label{eq:g}
\eeq
We turn now to the values to be associated to $\sigma_{\pi N}$, $y$ and $r$.

\subsection{Pion--nucleon sigma term.}
\label{sez:sigma}
The quantity $\sigma_{\pi N}$ may be deduced phenomenologically from measurements of the
pion--nucleon scattering; however, its derivation from the experimental 
data is rather involved. The customary procedure is to go through the 
following steps (see, for instance, Ref.\cite{reya}):

i) By use of phase--shift analysis and dispersion relations, from the 
experimental data of low--energy pion--nucleon scattering, one derives 
the quantity
\beq
\Sigma_{CD}\equiv \Sigma(t=2 m_{\pi}^2)\equiv f_{\pi}^2 
T_{\pi N}(s=m^2_N, t=2 m_{\pi}^2),
\eeq
\noindent where $s$ and $t$ are standard Mandelstam variables, $m_{\pi}$ is the pion mass, $f_{\pi}$
is the pion--decay constant and $T_{\pi N}$ is the (Born--subtracted) 
pion--nucleon scattering 
amplitude, calculated at the so--called Cheng--Dashen point.

ii) Modulo terms of order
$\lsim$ 1 MeV, which may be safely neglected, one has 
\beq
\Sigma_{CD} \simeq \sigma_{\pi N}(t=2 m_{\pi}^2), 
\eeq
\noindent where $\sigma_{\pi N}(t)$ is the nucleon scalar form factor,
defined as
\beq
\sigma_{\pi N}(t=(p^{\prime}-p)^2)\equiv <N(p^{\prime})|\frac{m_u+m_d}{2}
(\bar uu+\bar dd) |N(p)>.
\eeq 

iii) The evolution of $\sigma_{\pi N}(t)$, as a function of the momentum
transfer from $t=2 m_{\pi}^2$ to $t=0$, provides the value of $\sigma_{\pi N}
\equiv \sigma_{\pi N}(t=0)$: 
\beq
\sigma_{\pi N}=\sigma_{\pi N}(t=2 m_{\pi}^2)-\Delta \sigma.
\eeq

Now, it turns out that the determinations of $\Sigma_{CD}$ and $\Delta 
\sigma$ suffer from sizeable uncertainties.

For $\Sigma_{CD}$, apart from older calculations, we have, for instance: 
$\Sigma_{CD}=$56$\div$ 72 MeV
\cite{koch1} and $\Sigma_{CD}=$59$\div$ 62 MeV \cite{gls,gls2}.
A value of $\Sigma_{CD}$ on the high side ($\simeq$ 72 MeV) 
is also favoured by a more recent evaluation \cite{sainio}.
Thus, one could tentatively consider the range 
\beq \label{eq:rangesigmacd}
\Sigma_{CD}=56\div 72 \;\;{\rm MeV}.
\eeq
Calculations of $\Delta \sigma$ by dispersion relation techniques 
provide \cite{gls2} (see also \cite{bkm}): 
\beq
\Delta\sigma$=15.2$\pm 0.4 \;\;{\rm MeV} \label{eq:deltasigma}.
\eeq 
These values are much larger than the ones obtained with chiral 
perturbation theory at leading order $\Delta\sigma\simeq$ 7.5 MeV\cite{gl} 
(see Ref.\cite{gasser} for
a possible explanation of this discrepancy).
Furthermore, it has to be noted that $\Delta\sigma$ as deduced from lattice calculations\cite{dll}
is given by $\Delta\sigma$=6.6$\pm$0.6 MeV, thus by values sizably smaller 
than those of Eq.(\ref{eq:deltasigma}).

Combining the range for $\Sigma_{CD}$ in Eq.(\ref{eq:rangesigmacd}) and the value $\Delta\Sigma$=15 MeV,
we obtain for $\sigma_{\pi N}$ the range 41 MeV$\lsim \sigma_{\pi N}\lsim$57 MeV. This
may be compared to the results derived in Ref.\cite{bum} using heavy quark chiral perturbation theory:
38 MeV$\lsim \sigma_{\pi N}\lsim$58 MeV and in Ref.\cite{fkou} by lattice calculations:
40 MeV$\lsim \sigma_{\pi N}\lsim$60 MeV. Thus we conclude that the value of $\sigma_{\pi N}$
is still considerably uncertain; the previous results only indicate some 
convergence towards the range
$40 \;{\rm MeV}\lsim \sigma_{\pi N}\lsim 60 \;{\rm MeV}$,
 with an upper extreme which might be even higher ($\simeq$ 65 MeV),
should one take the chiral perturbation theory result ($\Delta \sigma$ $\simeq$ 7 MeV),
instead of the dispersion--relation result (Eq.(\ref{eq:deltasigma})).
Finally, we wish to notice that recent results from higher order chiral 
perturbation calculations provide a large value for $\sigma_{\pi N}$: 
$\sigma_{\pi N}$ = 70 MeV \cite{fms}; furthermore, use of a new pion-nucleon 
phase-shift analysis 
\cite{said}, instead of the standard  one \cite{koch}  would make  
$\sigma_{\pi N}$  to 
rocket to a value larger than 200 MeV \cite{bm}.

\subsection{Strange--quark content of the nucleon.}
\label{sez:strange}
A standard way to evaluate the quantity $y$ defined in Eq.(\ref{eq:y}) is to express it in terms of 
$\sigma_{\pi N}$ and of the quantity $\sigma_{0}$ defined as
\beq
\sigma_{0}\equiv \frac{1}{2}(m_u+m_d) <N|\bar uu+\bar dd-2\bar ss|N>,
\eeq
\noindent i.e.
\beq
y=1-\frac{\sigma_{0}}{\sigma_{\pi N}}.
\eeq
Actually, $\sigma_{0}$ is a quantity related to the size of the SU(3) symmetry breaking and, 
as such, may be calculated either from the octect baryon masses: 
$\sigma_{0}\simeq$33 MeV 
\cite{tpcheng} or with chiral perturbation theory: $\sigma_{0}$=35$\pm$5 MeV \cite{gl},
$\sigma_{0}$=36$\pm$7 MeV \cite{bum}. For definiteness, we take
\beq
\sigma_{0}=30\div40 {\rm MeV}.
\eeq
Thus, we have, for instance: 0$\leq y\leq$ 0.25 for $\sigma_{\pi N}$=40 MeV,
 0.11$\leq y\leq$ 0.33 for $\sigma_{\pi N}$=45 MeV, 
 0.33$\leq y\leq$ 0.50 for $\sigma_{\pi N}$=60 MeV, and 
 0.38$\leq y\leq$ 0.54 for $\sigma_{\pi N}$=65 MeV. 

We wish to remark that, by general physical arguments, one would expect for 
$y$ a somewhat small value,
i.e. $y\simeq$ 0.2--0.3; however, apart from the results of the previous derivations, which 
allow values of $y$ up to $y \simeq$ 0.5, also lattice results seem to favour 
large values:
$y$=0.36$\pm$0.03 \cite{dll}, $y$=0.66$\pm$0.15 \cite{fkou} (even reducing 
this latter value by 
$\sim$ 35\% as suggested in \cite{dll}, one would still have 
$y\simeq$0.4--0.5).
  
\subsection{Mass ratio $r= 2 m_s/(m_u+m_d)$.}
\label{sez:ratio}

As we have seen at the beginning of this section, besides
$\sigma_{\pi N}$ and $y$ a third ingredient is necessary for the evaluation 
of $m_q \langle N|\bar{q} q |N \rangle$ in the case of the strange quark
and the heavy ones: the mass ratio $r=2 m_s/(m_u+m_d)$.

The standard derivation of this ratio is based on chiral perturbation theory.
Lowest order formulae (corrected for electromagnetic effects) which give the
mass ratios $m_u/m_d$ and $m_s/m_d$ in terms of the physical masses of the 
$K$ mesons, entail, for $m_s/(m_u+m_d)$, the canonical value\cite{weinberg}

\beq
r\simeq 26.
\eeq 
The inclusion of next--to--leading order contributions in the chiral expansion
lead to the determination\cite{leut}
\beq
r=24.4\pm 1.5.
\eeq
Use of mass ratios in the evaluation of $r$ makes these results independent
of the renormalization scale. However, the validity of the chiral perturbation
method relies on the hypothesis that the quark condensate is the leading order
parameter of the spontaneously broken symmetry.

Other methods, most notably QCD sum rules and lattice simulations of QCD, are
capable of providing evaluations of individual quark masses (not only of their
ratios),
though these derivations still suffer from large uncertainties.
For the $u$ and $d$ quark masses we can quote the following results derived
from QCD sum rules:

\beqarr
(m_u+m_d)(1 \;{\rm GeV})&=& 12.0\pm 2.5 \cite{bpdv}\label{eq:mupiumd}\\
(m_u+m_d)(1 \;{\rm GeV})&=& 15.5\pm 2.0 \cite{ddr}.
\eeqarr 

Even more spread are the values deduced with this method by various authors
for the s--quark mass \cite{cdps,jm,cdfns,cps,dps}. A combination of all these
derivations: $m_s(1 \;{\rm GeV})=170\pm 50 \; {\rm MeV}$ \cite{dps} evidentiates 
how large is the uncertainty in the estimate of $m_s$. For sake of comparison,
we can quote some values for $m_s$ as derived from lattice QCD:
$m_s(1 \;{\rm GeV})=155\pm 15 \; {\rm MeV}$ \cite{dps,l,g} and from the $\tau$
hadronic width: $m_s(1\; {\rm GeV})=193\pm 59 \; {\rm MeV}$\cite{pp} and
$m_s(1 \;{\rm GeV})=200\pm 70 \; {\rm MeV}$ \cite{ckp}.

If, following Ref.\cite{bpdv}, one uses
\beq
m_s(1\; {\rm GeV})=175\pm 25 \; {\rm MeV}
\eeq
\noindent (which combines the results of Refs.\cite{cdps,jm}) and takes
$m_u+m_d=12.0\pm 2.5$ (see Eq.(\ref{eq:mupiumd})), one obtains 
\beq
r=29 \pm 7.
\eeq 
This result may be considered as representative of the uncertainty currently
affecting the mass ratio $r$.

\subsection{Size of the Higgs--quark couplings.}

From Eqs.(\ref{eq:condlight},\ref{eq:condstrange},\ref{eq:condheavy}) and the 
ranges for 
$\sigma_{\pi N}$, $y$ and $r$ discussed in Sections 
\ref{sez:sigma}--\ref{sez:ratio}, we see that the quantities 
$m_q \langle N|\bar{q} q |N \rangle$'s are indeed affected by large
uncertainties.
In particular, the quantity $m_s \langle N|\bar{s} s |N \rangle$, 
is uncertain roughly by a
factor of 3-4. 
This has quite significant consequences on the size of
$\sigma_{\rm scalar}^{\rm (nucleon)}$, because $m_s <N|\bar s s|N>$ is the most 
important term among the $m_q <N|\bar q q|N>$'s \cite{GGR}, unless $\tan \beta$ is very
small (see Eqs. (\ref{eq:k}),(\ref{eq:i}),(\ref{eq:g})).  

The values of the $m_q \langle N|\bar{q} q |N \rangle$'s for a few sets of
values for $\sigma_{\pi N}$, $y$ and $r$ are given in Table I. The first three
sets are representative of values currently employed in the literature. In
particular the third set (denoted here as set 1) is the one used in our previous
papers \cite{noi1234,bbbdfps} on the analysis of the DAMA annual modulation data.

Set 2 and set 3 are representative sets of values which are 
employed here to illustrate: i) to which extent the size of 
$\sigma_{\rm scalar}^{\rm (nucleon)}$  may be increased, within 
the afore mentioned uncertainties, and ii) which are the ensuing
implications for the neutralino cosmological properties, 
when these are
derived from experimental data of WIMP direct detection.   
Thus, sets 2 and 3 are meant only to be two possible representative 
set of values, chosen within the current 
uncertainties and capable of providing large values of the quantity 
$m_s <N|\bar s s|N>$. Because of the correlations, previously 
discussed, 
among $\sigma_{\rm scalar}^{\rm (nucleon)}$, $\sigma_{ann}$ and 
$\Omega_\chi h^2$, sets 2 and 3 are expected to provide sizeable values 
for the neutralino relic abundance. 

A more systematic analysis of the implications for relic neutralinos 
of uncertainties in the Higgs--quark couplings will be feasible, 
only when a consistent field of variation for the correlated 
quantities $\sigma_{\pi N}$, $y$ and $r$ will emerge from a thorough 
and coherent QCD investigation of the problem.

\section{Results and conclusions}

Let us now turn to the presentation of our results. 
In Fig.1.a (1.b) we give the ratio of the cross--section 
$\sigma_{\rm scalar}^{\rm (nucleon)}$ calculated with set 2 (set 3), 
to $\sigma_{\rm scalar}^{\rm (nucleon)}$ calculated with set 1. We see 
that for $(\sigma_{\rm scalar}^{\rm (nucleon)})_{\rm set\; 1}$ in the 
range of Eq. (\ref{eq:ranges}) most configurations cluster around the 
values  
$(\sigma_{\rm scalar}^{\rm (nucleon)})_{\rm set\;2} /
(\sigma_{\rm scalar}^{\rm (nucleon)})_{\rm set\;1} \simeq 3$, 
$(\sigma_{\rm scalar}^{\rm (nucleon)})_{\rm set\;3} /
(\sigma_{\rm scalar}^{\rm (nucleon)})_{\rm set\;1} \simeq 5$.
Thus, we have a sizeable 
increase in the cross--section, when sets 2 or 3 are used instead of 
our set of reference (set 1), which was utilized in 
Refs. \cite{noi1234,bbbdfps}. 

Fig.2 displays the plot of $\sigma_{\rm scalar}^{\rm (nucleon)}$ versus 
$\Omega_\chi h^2$, in case of set 1 (Fig. 2a) and of set 2 (Fig. 2b). 
The two horizontal dashed lines delimit the range of the neutralino-nucleon
cross section defined in Eq.(\ref{eq:ranges}).
The solid vertical lines  delimit the cosmologically 
interesting range $0.01 \leq \Omega_{\chi} h^2 \leq 0.7$.  The two vertical
dashed lines delimit the range: 
$0.02 \leq \Omega_{\chi} h^2 \leq 0.2$, which represents a
particularly appealing 
interval, according to the most recent observations and analyses
\cite{omegamatter,hubble}.
 The overall shape of these scatter plots reflects the 
anticorrelation between 
$\sigma_{\rm scalar}^{\rm (nucleon)}$ and $\Omega_\chi h^2$. 
 The most relevant 
feature of these  plots is the  boundary on the 
top--right  side. It provides the maximal values allowed for the 
relic density, for neutralino--nucleon cross--sections in the range of 
Eq. (\ref{eq:ranges}). Notice how the boundary extends  
into  the region of cosmological interest more markedly in case of 
set 2 than in case of set 1.  This is a first manifestation of the 
fact that the DAMA annual modulation effect is compatible with 
a neutralino of cosmological interest, to an extent which is even 
larger than the one singled out in our previous papers 
\cite{noi1234,bbbdfps}, where only 
the representative set 1 was employed.
(Notice that in the present analysis, although the overall
range of variation of the susy parameters is the same as in 
Ref.\cite{noi1234,bbbdfps}, we have optimized the numerical
scanning of the parameter space in order to increase the 
density of configurations which fall in the region of main
interest.)

These cosmological properties are further displayed
in Fig. 3, which depicts the scatter plots of $\rho_{\chi}$ versus 
$\Omega_\chi h^2$. Here the two horizontal lines delimit the physical 
region for the total local density of non--baryonic dark matter:
0.1 GeV cm$^{-3} \leq \rho_{\chi} \leq 0.7$ GeV cm$^{-3}$;
the two slant 
dot--dashed lines delimit the band, where linear rescaling procedure 
for the local density is usually applied \cite{noi1234,bbbdfps}. 
At variance with the previous scatter plots of Figs. 1-2, which 
refer to a generic scanning of the supersymmetric parameter 
space, constrained only by accelerator bounds, as discussed 
in Refs. \cite{noi1234,bbbdfps}, the scatter plots of Fig. 3  
display the susy configurations singled out by the DAMA annual 
modulation data. 
These plots are obtained with the following procedure: 

i) $\rho_{\chi}$ is evaluated as 
$\rho_{\chi} = 
[\rho_{\chi} \sigma_{\rm scalar}^{\rm (nucleon)}]_{R_m}/
\sigma_{\rm scalar}^{\rm (nucleon)}$, where 
$[\rho_{\chi} \sigma_{\rm scalar}^{\rm (nucleon)}]_{R_m}$ denotes the set 
of experimental values of $\rho_{\chi} \sigma_{\rm scalar}^{\rm (nucleon)}$ 
inside the DAMA annual modulation region $R_m$ and
$\sigma_{\rm scalar}^{\rm (nucleon)}$ is calculated with Eq.(\ref{eq:sigma}).

ii) To each value of $\rho_{\chi}$, which then pertains to a specific 
supersymmetric configurations, one associates the corresponding 
value of $\Omega_\chi h^2$, calculated as in Ref. \cite{noiomega}. 

Therefore, with this procedure, we determine the values of $\rho_\chi$
which, for each calculated $\sigma_{\rm scalar}^{\rm (nucleon)}$,
satisfy the DAMA annual modulation data.

Fig. 3 shows that the set of supersymmetric configurations selected 
by the DAMA data has a significant overlap with the region of main 
cosmological interest:  $\Omega_\chi h^2 \gsim 0.02$ and
0.1 GeV cm$^{-3} \leq \rho_{\chi} \leq 0.7$ GeV cm$^{-3}$. 
The extent of this overlap is increasingly larger for set 2 
and set 3 of Table I. By way of example, for set 3 one has that, 
at $\rho_{\chi} = 0.3$ GeV cm$^{-3}$, $\Omega_\chi h^2$ may reach the value 
0.3.  Therefore, these results reinforce our conclusions of  
Ref. \cite{noi1234,bbbdfps}, {\it i.e.} that the DAMA annual 
modulation data are compatible with a neutralino as a major component 
of dark matter, on the average in the Universe and in our Galaxy. 
 
The same figure shows that different situations are also possible.
Specifically, for configurations which fall inside 
the band delimited by the slant dot--dashed lines,
the neutralino would provide only a fraction of the cold dark 
matter both at the level of local density and at the level of the 
average $\Omega$, a situation which would be possible, for instance,
if the neutralino is not the unique cold dark matter particle
component. On the other hand, configurations above the 
upper dot--dashed line and below the upper horizontal solid line 
would imply a stronger clustering of neutralinos in our halo as 
compared to their average distribution in the Universe. This
situation may be considered unlikely, since in this case
neutralinos could fulfill the experimental range for 
$\rho_\chi$, but they would contribute only a small fraction to
the cosmological cold dark matter content.
Finally, configurations above the upper horizontal line are
incompatible with the upper limit on the local density of dark
matter in our Galaxy and must be disregarded.

Finally, we wish to add the following comments:

i) To show the effect of the current uncertainties in the 
Higgs--quark couplings, a few representative sets of values for the 
relevant quantities ($\sigma_{\pi N}$, $y$ and $r$)
were employed, as illustration of the implications 
for relic neutralinos. 
We have explicitly discussed situations which allow sizeable
values of the neutralino relic abundance. It is clear that,
within the same uncertainties, other sets of values of
the relevant parameters would entail relic neutralinos of
much less appealing cosmological interest.
A more precise evaluation of the effects implied 
by the mentioned uncertainties will require a significant 
breakthrough in the understanding of the QCD and of other 
hadron aspects involved in the 
problem. What is lacking at present is consistency among the 
various determinations in $\sigma_{\pi N}$, $y$ and $r$. 

ii) The relevance of the implication of the mentioned uncertainties 
in converting information from measurements of neutralino detection 
to neutralino cosmological properties, discussed here for experiments 
of direct detection, applies also to indirect experiments at neutrino 
telescopes, when one looks at upgoing muons from the center of the 
Earth.  In fact, in this case the size of the expected signal 
depends on the neutralino capture rates by the Earth, and then 
 mainly on $\rho_{\chi} \sigma_{\rm scalar}^{\rm (nucleon)}$. 

iii) A number of experimental means, meant to search for the 
presence of 
relic neutralinos in our galaxy, look for signals (antiprotons,
antideuterons, positrons, diffuse gamma--rays, gamma lines) \cite{bf}  
originated by neutralino--neutralino annihilation in the halo. Thus, 
the relevant cross--section involved in this case is $\sigma_{ann}$. 
Therefore, a word of caution is in order here, for the case when 
the expectations for these indirect signals are based on 
supersymmetric configurations derived from direct detection data. 
In the estimate of these indirect signals, one should include the 
uncertainties discussed in the present paper.  
 
\acknowledgements
This work was partially supported by the Spanish DGICYT under grant number 
PB95--1077, by the TMR network grant ERBFMRXCT960090 of the European
Union and by the Research Grants of the Italian Ministero
dell'Universita' e della Ricerca Scientifica e Tecnologica.

\newpage
\begin{table}[h]
\begin{center}
\begin{tabular}{|c|c|c|c|c|c|cc|}   \hline
 $\sigma_{\pi N}$ & $y$  & $r$ & $m_{q_l}<N|\bar{q_l}q_l|N>$ & $m_s<N|\bar{s}s|N>$ 
& $m_h<N|\bar{h}h|N>$ & & Ref.  \\ 
(MeV)   &   &  & (MeV)  & (MeV)  & (MeV)  & &  \\ \hline
45 &0.28  & 25 & 23 & 158 & 55 & &\protect\cite{don} \\ \hline
45   &    &  & 27 & 131 & 56  & &\protect\cite{jkg,bg} \\ \hline
45 & 0.33 & 29 & 23 & 215 & 50 & set 1 &\protect\cite{bbbdfps,noi1234,bbl} \\ \hline
60 & 0.50 & 29 & 30 & 435 & 33 & set 2 & \\ \hline
65 & 0.50 & 36 & 33 & 585 & 21 & set 3 & \\ \hline
\end{tabular}
\end{center}
\caption{Values of the matrix elements $\langle N|\bar{q} q |N \rangle$
of the quark scalar densities in the nucleon times the quark masses $m_q$,
for a few sets of values of the pion--nucleon sigma term $\sigma_{\pi N}$, 
the fractional strange--quark content of the nucleon $y$ and the quark mass
ratio $r=2m_s/(m_u+m_d)$. $q_l$ stands for light quarks, $s$ is
the strange quark and $h=c,b,t$ denotes heavy quarks. For the light
quarks, we have defined 
$m_{q_l}<N|\bar{q_l}q_l|N>$ $\equiv$ 
$\frac{1}{2}[m_u <N|\bar u u|N> + m_d <N|\bar d d|N>]$.
}
\end{table}

\newpage
\begin{center}
{\Large FIGURE CAPTIONS}
\end{center}
\vspace{1cm}

FIG. 1a.
Ratio of the neutralino--nucleon scalar cross--section 
$\left(\sigma_{\rm scalar}^{(\rm nucleon)}\right)_{\rm set \; 2}$, 
calculated with the parameters of set 2 defined in Table I, 
to 
$\left(\sigma_{\rm scalar}^{(\rm nucleon)}\right)_{\rm set \; 1}$
calculated with set 1, as a 
function of $\left(\sigma_{\rm scalar}^{(\rm nucleon)}\right)_{\rm set \; 1}$.

\vspace{0.7cm}

FIG. 1b.
The same as in Fig.1a, with set 3 instead of set 2.

\vspace{0.7cm}

FIG. 2a.
Neutralino--nucleon scalar cross--section 
$\sigma_{\rm scalar}^{(\rm nucleon)}$, calculated using  
the parameters of set 1, as a function of the neutralino 
relic abundance $\Omega_{\chi} h^2$. 
The two horizontal dashed lines delimit the range of the neutralino-nucleon
cross section defined in Eq.(\ref{eq:ranges}).
The two solid vertical lines delimit the interval of cosmological
interest. The two vertical dashed lines delimit the preferred band
for cold dark matter.
The shaded region is cosmologically excluded on the basis of present
limits on the age of the Universe.

\vspace{0.7cm}

FIG. 2b.
The same as in Fig.2a, for 
the parameters of set 2.

\vspace{0.7cm}

FIG. 3a.
Neutralino local density $\rho_{\chi}$ derived by
requiring that $\rho_{\chi} \sigma_{\rm scalar}^{\rm (nucleon)}$ 
falls inside the experimental DAMA region $R_m$,
plotted against the neutralino relic abundance $\Omega_{\chi} h^2$. 
The quantity $\sigma_{\rm scalar}^{\rm (nucleon)}$ is calculated with
the parameters of set 1.
The two horizontal lines delimit the physical range for the local 
density of non-baryonic dark matter. The two solid vertical lines delimit 
the interval of $\Omega_{\chi} h^2$ of cosmological interest. The 
two vertical dashed lines delimit the  preferred band for cold 
dark matter. The two slant  dot--dashed lines delimit the 
band where linear rescaling procedure is usually applied. 
The shaded region is cosmologically excluded on the basis of present
limits on the age of the Universe.
Different symbols identify different neutralino
compositions: circles stand for a higgsino, crosses for a gaugino
and dots for a mixed neutralino. 

\vspace{0.7cm}

FIG. 3b.
The same as in Fig.3a, for 
the parameters of set 2.

\vspace{0.7cm}

FIG. 3c.
The same as in Fig.3a, for 
the parameters of set 3.

\vspace{0.7cm}

\newpage
\begin{figure}[t]
\hbox{
\psfig{figure=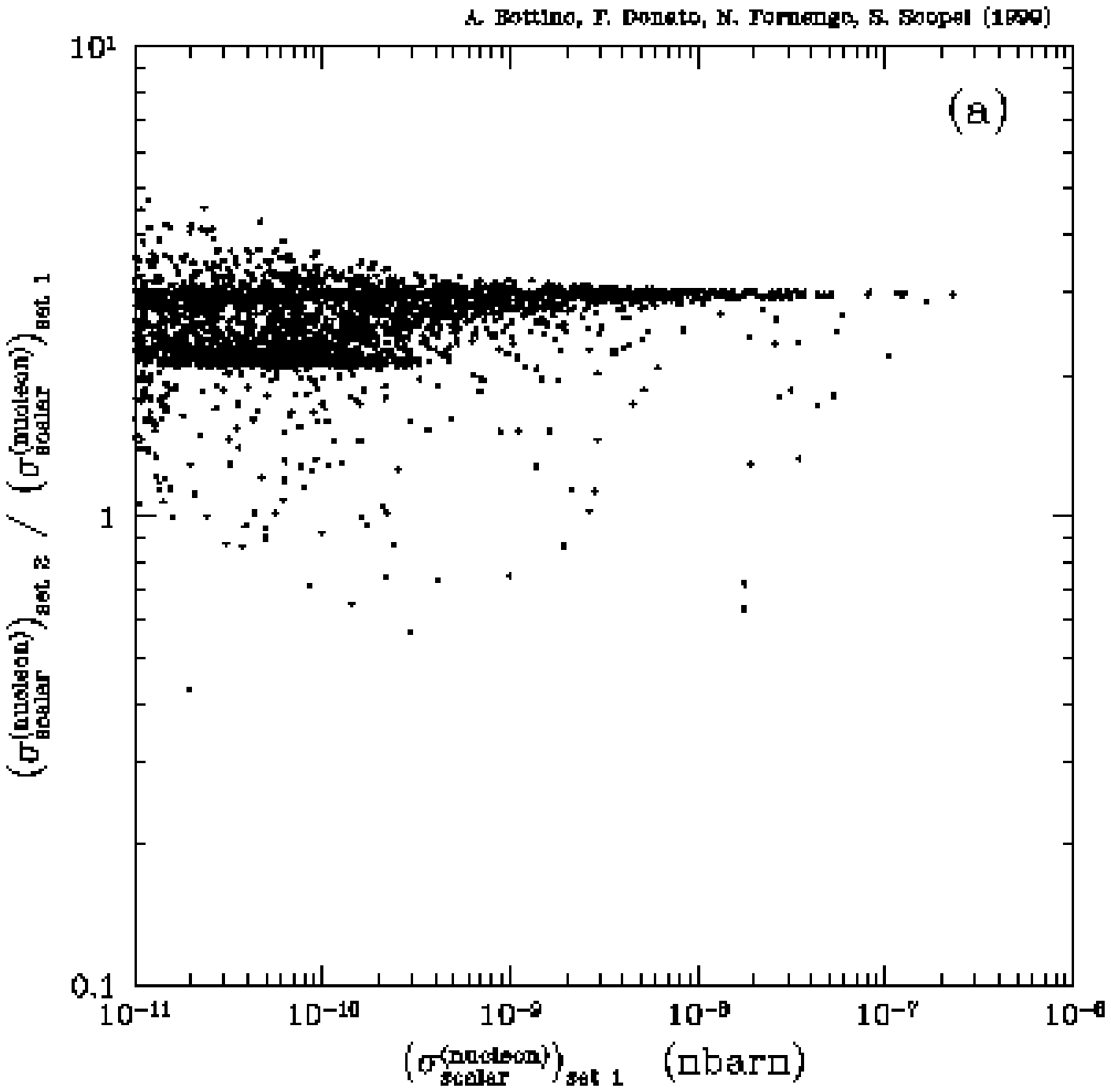,width=8.2in,bbllx=40bp,bblly=160bp,bburx=700bp,bbury=660bp,clip=}

}
{
FIG.1a - Ratio of the neutralino--nucleon scalar cross--section 
$\left(\sigma_{\rm scalar}^{(\rm nucleon)}\right)_{\rm set \; 2}$, 
calculated with the parameters of set 2 defined in Table I, 
to 
$\left(\sigma_{\rm scalar}^{(\rm nucleon)}\right)_{\rm set \; 1}$
calculated with set 1, as a 
function of $\left(\sigma_{\rm scalar}^{(\rm nucleon)}\right)_{\rm set \; 1}$.
}
\end{figure}

\newpage
\begin{figure}[t]
\hbox{
\psfig{figure=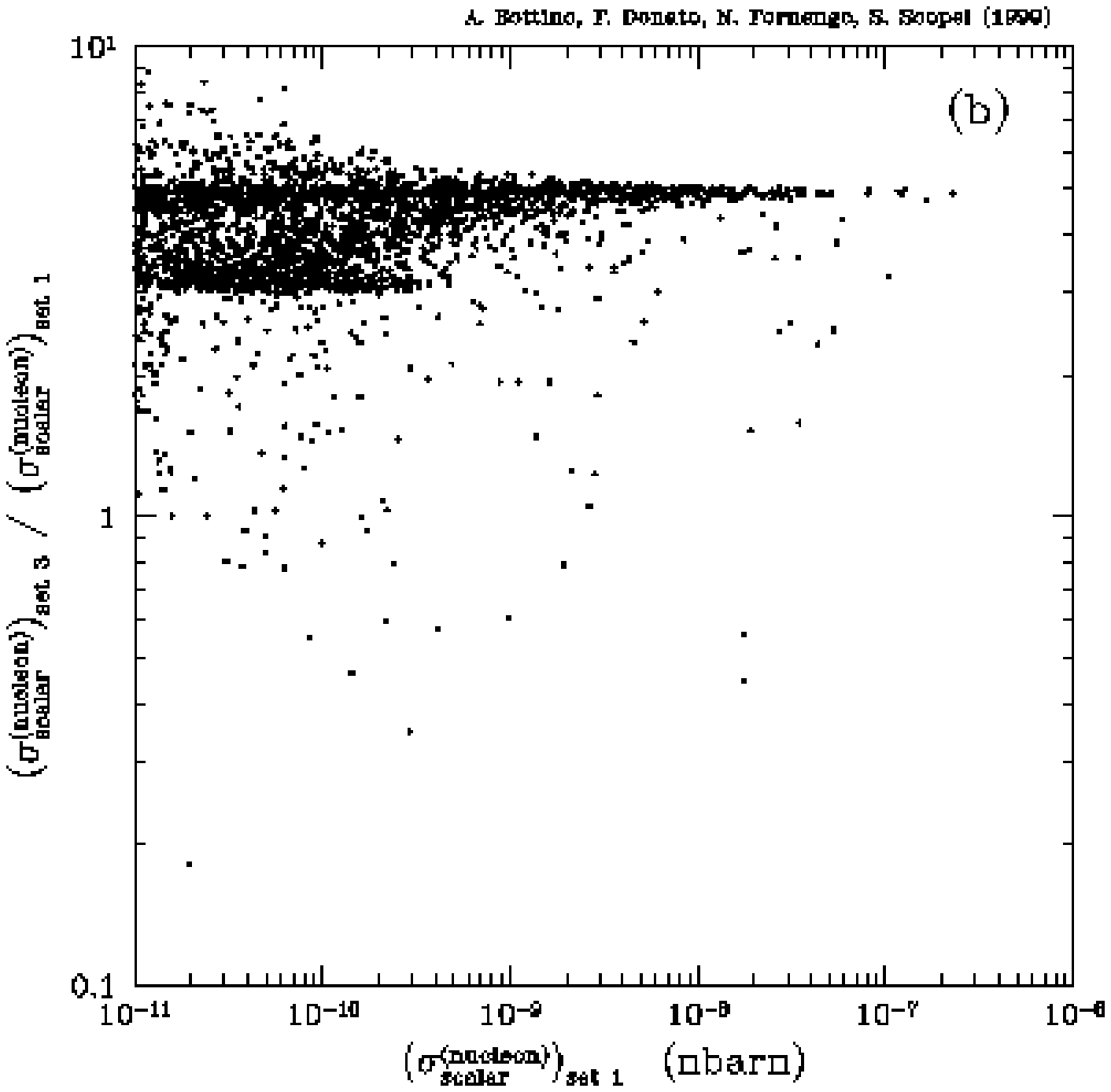,width=8.2in,bbllx=40bp,bblly=160bp,bburx=700bp,bbury=660bp,clip=}
}
{
FIG.1b - The same as in Fig.1a, with set 3 instead of set 2.}
\end{figure}

\newpage
\begin{figure}[t]
\hbox{
\psfig{figure=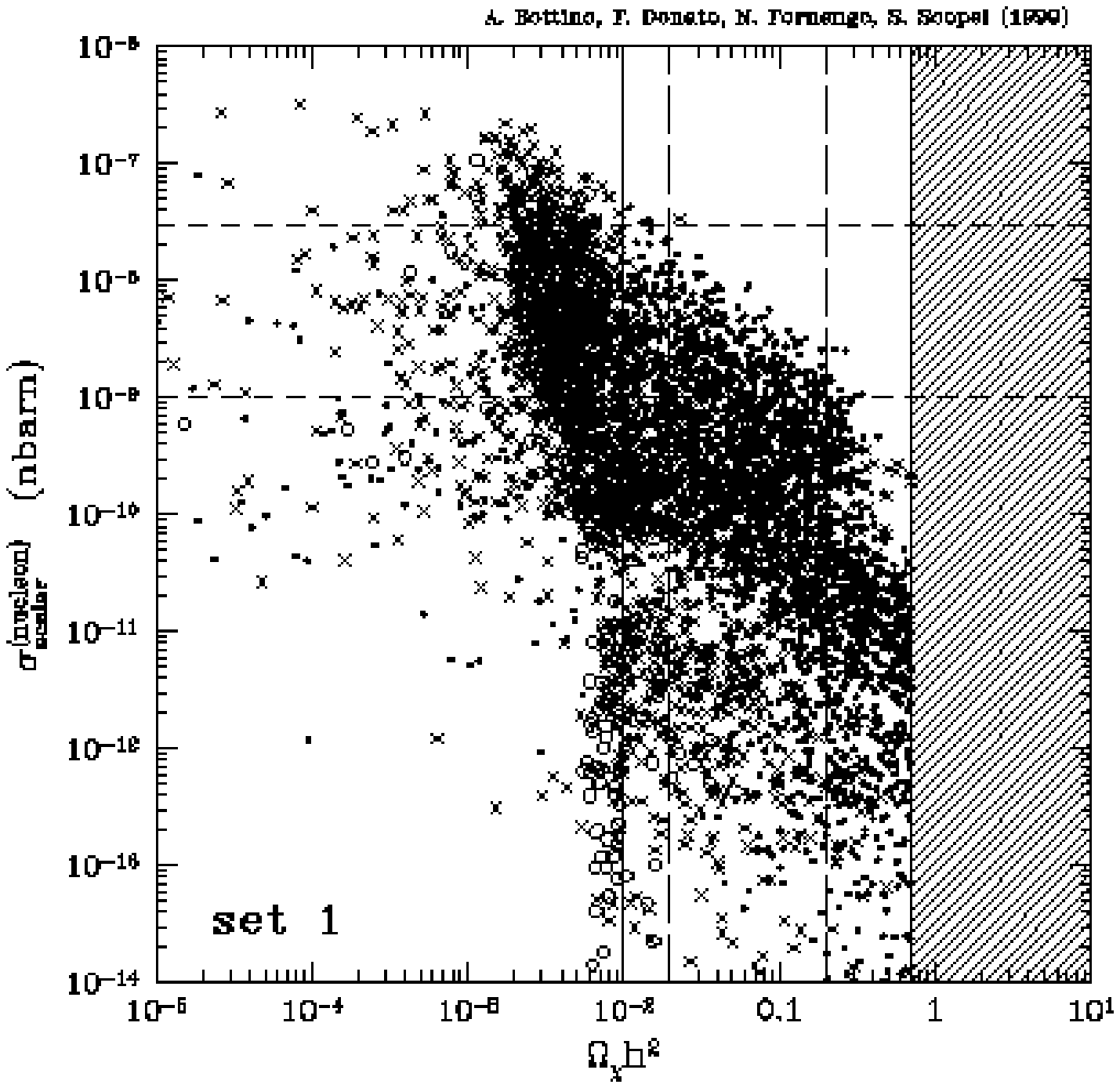,width=8.2in,bbllx=40bp,bblly=160bp,bburx=700bp,bbury=660bp,clip=}
}
{
FIG.2a - Neutralino--nucleon scalar cross--section 
$\sigma_{\rm scalar}^{(\rm nucleon)}$, calculated using  
the parameters of set 1, as a function of the neutralino 
relic abundance $\Omega_{\chi} h^2$. 
The two horizontal dashed lines delimit the range of the neutralino-nucleon
cross section defined in Eq.(\ref{eq:ranges}).
The two solid vertical lines delimit the interval of cosmological
interest. The two vertical dashed lines delimit the preferred band
for cold dark matter.
The shaded region is cosmologically excluded on the basis of present
limits on the age of the Universe.
}
\end{figure}

\newpage
\begin{figure}[t]
\hbox{
\psfig{figure=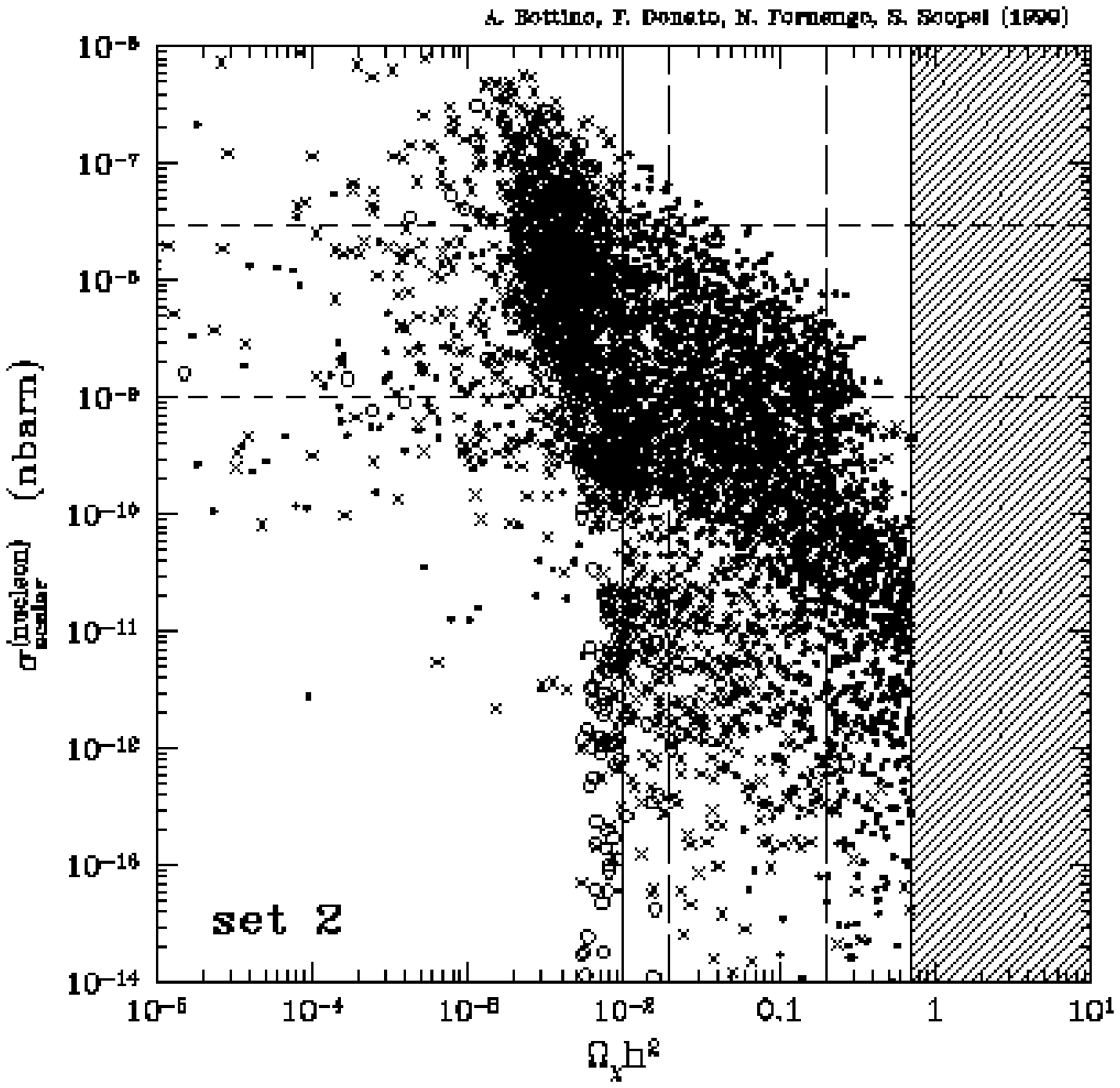,width=8.2in,bbllx=40bp,bblly=160bp,bburx=700bp,bbury=660bp,clip=}
}
{
FIG.2b - The same as in Fig.2a, for 
the parameters of set 2.
}
\end{figure}

\newpage
\begin{figure}[t]
\hbox{
\psfig{figure=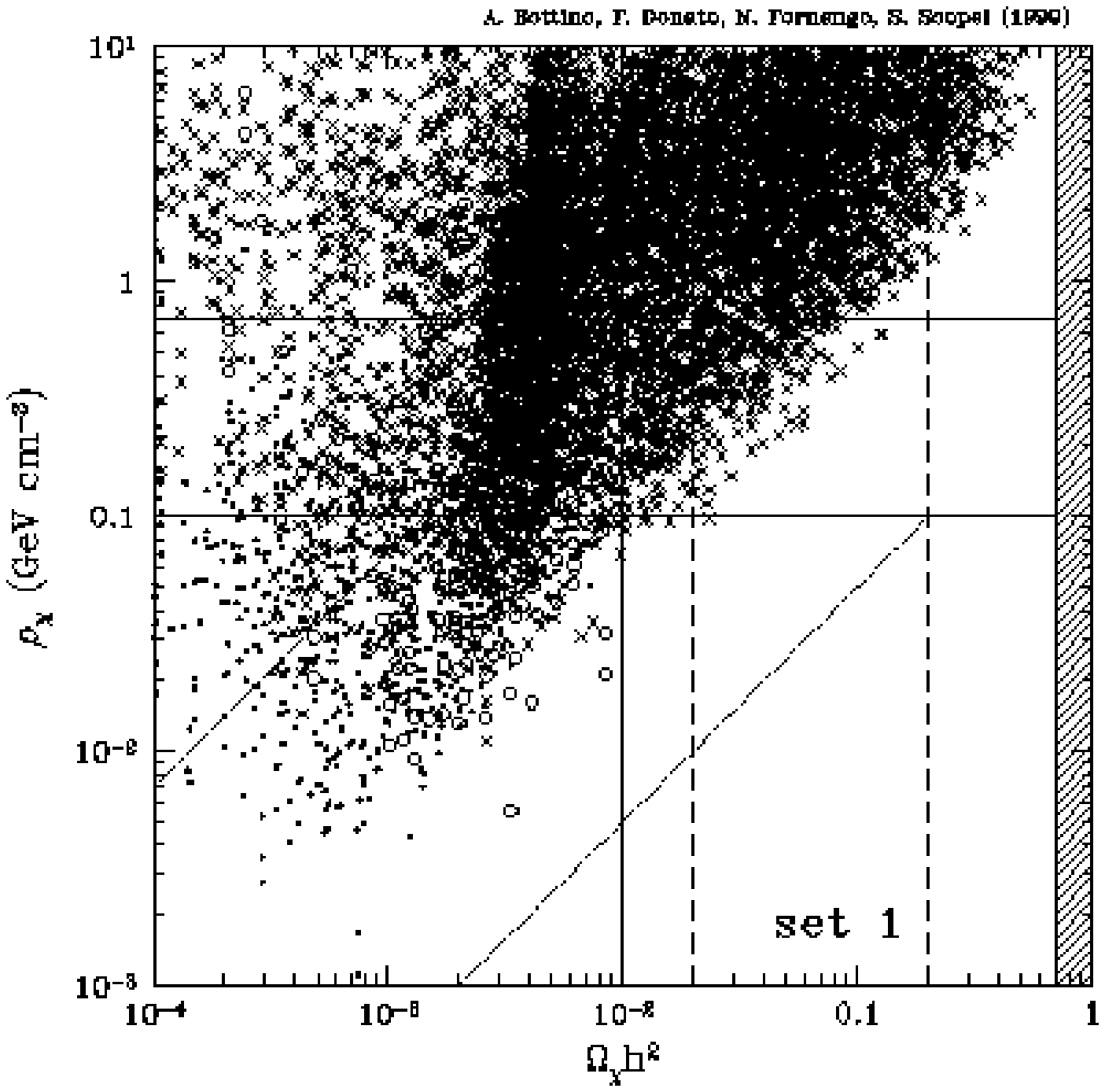,width=8.2in,bbllx=40bp,bblly=160bp,bburx=700bp,bbury=660bp,clip=}
}
{
FIG.3a - Neutralino local density $\rho_{\chi}$ derived by
requiring that $\rho_{\chi} \sigma_{\rm scalar}^{\rm (nucleon)}$ 
falls inside the experimental DAMA region $R_m$,
plotted against the neutralino relic abundance $\Omega_{\chi} h^2$. 
The quantity $\sigma_{\rm scalar}^{\rm (nucleon)}$ is calculated with
the parameters of set 1.
The two horizontal lines delimit the physical range for the local 
density of non-baryonic dark matter. The two solid vertical lines delimit 
the interval of $\Omega_{\chi} h^2$ of cosmological interest. The 
two vertical dashed lines delimit the  preferred band for cold 
dark matter. The two slant  dot--dashed lines delimit the 
band where linear rescaling procedure is usually applied. 
The shaded region is cosmologically excluded on the basis of present
limits on the age of the Universe.
Different symbols identify different neutralino
compositions: circles stand for a higgsino, crosses for a gaugino
and dots for a mixed neutralino.
}
\end{figure}

\newpage
\begin{figure}[t]
\hbox{
\psfig{figure=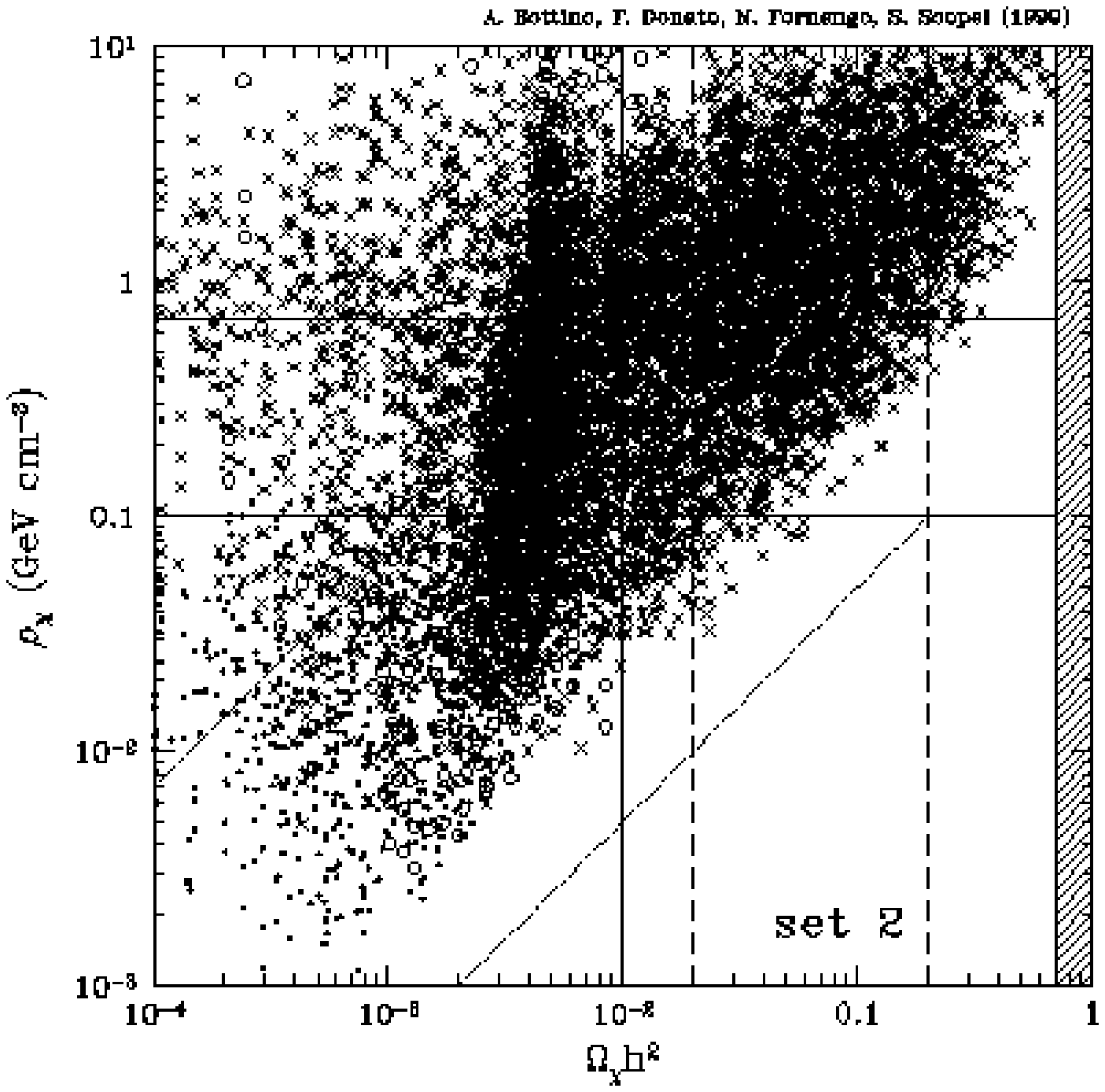,width=8.2in,bbllx=40bp,bblly=160bp,bburx=700bp,bbury=660bp,clip=}
}
{
FIG.3b - The same as in Fig.3a, for 
the parameters of set 2.
}
\end{figure}

\newpage
\begin{figure}[t]
\hbox{
\psfig{figure=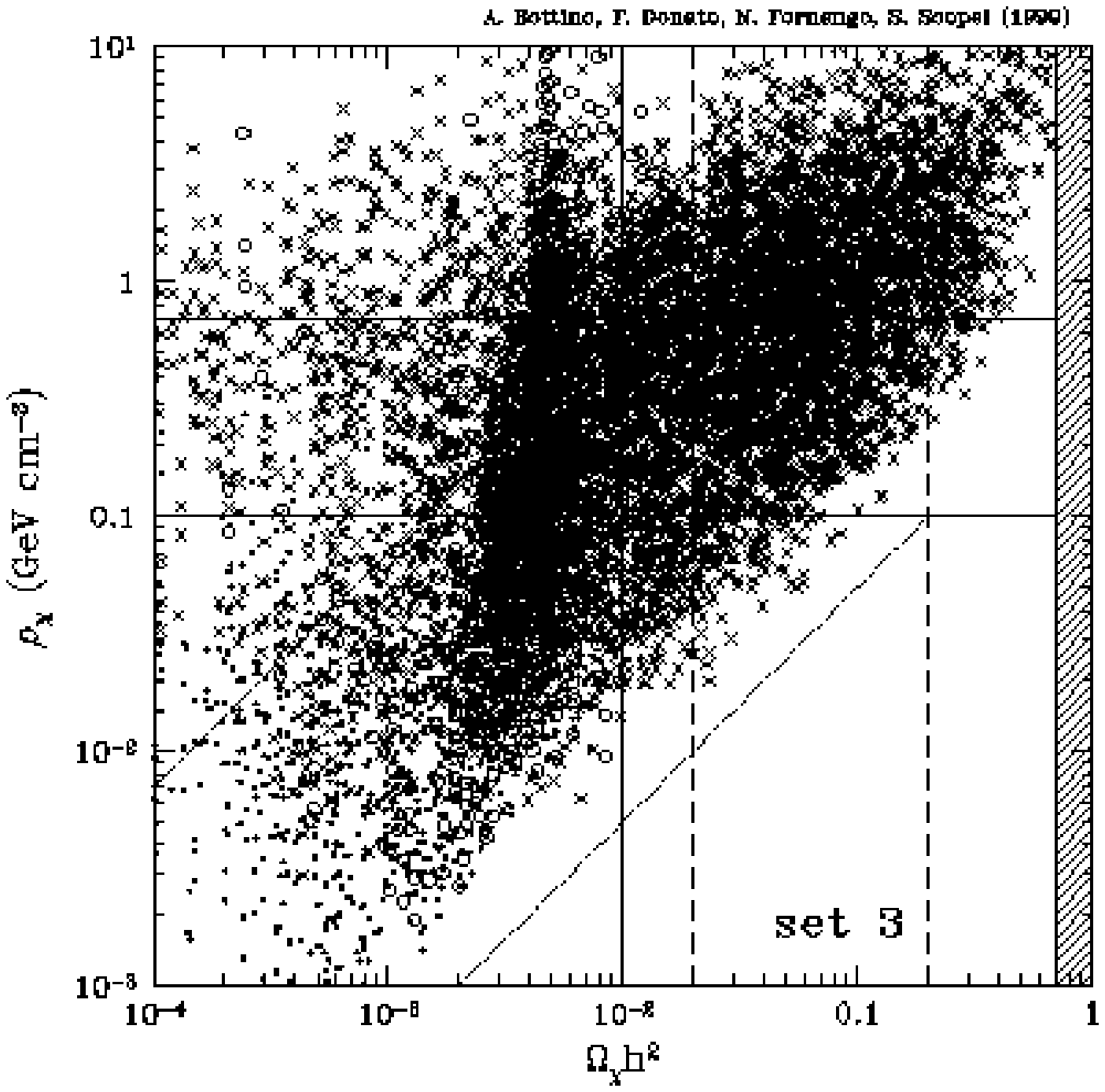,width=8.2in,bbllx=40bp,bblly=160bp,bburx=700bp,bbury=660bp,clip=}
}
{
FIG.3c - The same as in Fig.3a, for 
the parameters of set 3.
}
\end{figure}


\begin{thebibliography}{10}
\bibitem{den} Although a number of considerations presented in Sect. I apply to a
wide class of WIMPs, the remain of the paper mainly refer to the neutralino. 
To simplify notations, the WIMP local density is everywhere denoted by 
$\rho_{\chi}$, even
when the identification WIMP$\equiv \chi$ is not implied.   The same for the
WIMP mass, which is here always denoted by $m_{\chi}$.
\end{thebibliography}
\end{document}